\documentclass[fleqn,usenatbib]{mnras}

\usepackage{newtxtext,newtxmath}

\usepackage[T1]{fontenc}
\usepackage{ae,aecompl}

\usepackage{graphicx}	
\usepackage{amsmath}	
\usepackage{amssymb}	
\usepackage{epstopdf}
\usepackage{multirow}

\title[Connecting HMXB $Z$-dependence and $z$-evolution]{Connecting the metallicity dependence and redshift evolution of high-mass X-ray binaries}

\author[F. M. Fornasini et al.]{
Francesca M. Fornasini,$^{1}$\thanks{E-mail: francesca.fornasini@cfa.harvard.edu}
Francesca Civano,$^{1}$
and Hyewon Suh$^{2}$
\\
$^{1}$Center for Astrophysics | Harvard \& Smithsonian, 60 Garden Street, Cambridge, MA 02138, USA\\
$^{2}$Subaru Telescope, National Astronomical Observatory of Japan (NAOJ), 650 North A'ohoku place, Hilo, HI 96720, USA\\
}

\date{Accepted XXX. Received YYY; in original form ZZZ}

\pubyear{2019}

\newcommand\T{\rule{0pt}{2.6ex}}       
\newcommand\B{\rule[-1.2ex]{0pt}{0pt}} 
\defcitealias{mineo12}{M12}
\defcitealias{lehmer10}{L10}
\defcitealias{lehmer16}{L16}
\defcitealias{lehmer19}{L19}
\defcitealias{fragos13b}{F13}
\defcitealias{madau17}{M17}
\defcitealias{aird17}{A17}
\defcitealias{brorby16}{B16}
\defcitealias{fornasini19}{F19}
\defcitealias{fornasini18}{F18}

\begin{document}
\label{firstpage}
\pagerange{\pageref{firstpage}--\pageref{lastpage}}
\maketitle

\begin{abstract}
The integrated X-ray luminosity ($L_{\mathrm{X}}$) of high-mass X-ray binaries (HMXBs) in a galaxy is correlated with its star formation rate (SFR), and the normalization of this correlation increases with redshift.  Population synthesis models suggest that the redshift evolution of $L_{\mathrm{X}}$/SFR is driven by the metallicity ($Z$) dependence of HMXBs, and the first direct evidence of this connection was recently presented using galaxies at $z\sim2$.  To confirm this result with more robust measurements and better constrain the $L_{\mathrm{X}}$-SFR-$Z$ relation, we have studied the $Z$ dependence of $L_{\mathrm{X}}$/SFR at lower redshifts.  Using samples of star-forming galaxies at $z=0.1-0.9$ with optical spectra from the hCOSMOS and zCOSMOS surveys, we stacked \textit{Chandra} data from the COSMOS Legacy survey to measure the average $L_{\mathrm{X}}$/SFR as a function of $Z$ in three redshift ranges: $z=0.1-0.25$, $0.25-0.4$, and $0.5-0.9$.  We find no significant variation of the $L_{\mathrm{X}}$-SFR-$Z$ relation with redshift.  Our results provide further evidence that the $Z$ dependence of HMXBs is responsible for the redshift evolution of $L_{\mathrm{X}}$/SFR.  Combining all available $z>0$ measurements together, we derive a best-fitting $L_{\mathrm{X}}$-SFR-$Z$ relation and assess how different population synthesis models describe the data.  These results provide the strongest constraints to date on the $L_{\mathrm{X}}$-SFR-$Z$ relation in the range of $8.0<$12+log(O/H)$<9.0$.
\end{abstract}



\section{Introduction}
High-mass X-ray binaries (HMXBs), which consist of a neutron star or black hole accreting material from a stellar companion with $M_*>10M_{\odot}$, are young stellar systems with ages of $\sim5-50$ Myr (\citealt{iben95}; \citealt{bodaghee12c}; \citealt{antoniou16}).  As a consequence, the number of HMXBs in a galaxy and their integrated X-ray luminosity ($L_{\mathrm{X}}$) is correlated with a galaxy's star formation rate (SFR; \citealt{ranalli03}; \citealt{grimm03}; \citealt{persic04}; \citealt{gilfanov04}; \citealt{lehmer10}; \citealt{mineo12} hereafter \citetalias{mineo12}; \citealt{lehmer19} hereafter \citetalias{lehmer19}).  This correlation exhibits large scatter of $\approx0.4$ dex \citep{mineo12} and its normalization increases with redshift (\citealt{basu13a}; \citealt{lehmer16}; \citealt{aird17}).  \par
Binary population synthesis models have suggested that the redshift evolution and part of the scatter of the $L_{\mathrm{X}}^{\mathrm{HMXB}}$-SFR relation may be driven by the metallicity ($Z$) dependence of HMXB evolution (\citealt{dray06}; \citealt{linden10}; \citealt{fragos13a}; \citealt{fragos13b} hereafter \citetalias{fragos13b}; \citealt{madau17}).  In these models, the $Z$-dependence of HMXBs arises from the fact that higher-$Z$ stars have more powerful radiatively-driven winds.  As a result, high-$Z$ stars lose more mass prior to exploding as supernovae and high-$Z$ binaries lose more angular momentum compared to their low-$Z$ counterparts.  Thus, these studies predict that lower-$Z$ HMXB populations should host more massive compact objects and more Roche-lobe overflow systems, resulting in higher accretion rates and higher $L_{\mathrm{X}}$ on average.  \par
In general agreement with these predictions, it has been observed that local low-$Z$ dwarf galaxies do host a larger number of luminous HMXBs at fixed SFR compared to high-$Z$ galaxies (\citealt{mapelli11}; \citealt{kaaret11}; \citealt{prestwich13}; \citealt{basu13b}; \citealt{brorby14}; \citealt{douna15}; \citealt{basu16}; \citetalias{lehmer19}; \citealt{ponnada20}).  Expanding on work by \citet{douna15}, \citet{brorby16} (hereafter \citetalias{brorby16}) used a sample of 49 nearby galaxies to measure the anti-correlation between $L_{\mathrm{X}}$/SFR and $Z$, and found decent agreement with the theoretical expectations of \citet{fragos13b}.  It is unclear whether the excess number of bright HMXBs in low-$Z$ galaxies can be explained by an increase in the normalization of the HMXB luminosity function from higher-$Z$ galaxies (\citealt{brorby14}; \citealt{ponnada20}), whether a shallower bright end slope to the HMXB luminosity function is also required \citep{basu16}, or whether the HMXB luminosity function varies with $Z$ in more complicated ways \citepalias{lehmer19}.  While it has been shown that significant scatter in the $L_{\mathrm{X}}$-SFR relation can be introduced at low SFR as a result of incomplete Poissonian sampling of the HMXB luminosity function (\citealt{gilfanov04a}; \citealt{justham12}; \citetalias{lehmer19}), the results of the aforementioned studies clearly indicate that $Z$ variations in galaxy samples can introduce scatter as well.  \par
Recently, \citet{fornasini19} (hereafter \citetalias{fornasini19}) presented the first direct evidence that the redshift evolution of HMXBs can be attributed to the $Z$ dependence of $L_{\mathrm{X}}^{\mathrm{HMXB}}$/SFR, as hypothesized by previous studies (\citealt{basu13a}; \citealt{fragos13a}; \citealt{lehmer16}). By stacking the X-ray data of $z\sim2$ galaxies grouped into $Z$ bins, \citetalias{fornasini19} found that $L_{\mathrm{X}}$/SFR is anti-correlated with $Z$ at $z\sim2$, and that, for a given $Z$, the $L_{\mathrm{X}}$/SFR values of HMXB-dominated galaxies at this redshift are consistent with the local $L_{\mathrm{X}}$-SFR-$Z$ relation measured by \citetalias{brorby16}.  Thus, on average, $z\sim2$ galaxies have higher $L_{\mathrm{X}}$/SFR values than $z=0$ galaxies of similar stellar mass ($M_*$) because the average $Z$ is lower at $z\sim2$. \par
However, the observed $L_{\mathrm{X}}$/SFR-$Z$ anti-correlation at $z\sim2$ is significant only at 97\% confidence \citepalias{fornasini19}.  Furthermore, it is possible that measurements of the local $L_{\mathrm{X}}$-SFR-$Z$ relation may be biased due to selection effects (\citealt{douna15}; \citetalias{brorby16}), and therefore may not provide a representative benchmark for $z=0$ HMXBs.  Thus, in order to make a more robust conclusion about the extent to which the $Z$ dependence of HMXBs drives the redshift evolution of $L_{\mathrm{X}}$/SFR, it is important to study the $Z$ dependence of HMXBs over a broad redshift range. \par
We present a study of the $Z$ dependence of HMXBs in three redshift intervals: $z=0.1-0.25$, $0.25-0.4$, and $0.5-0.9$. The goal of this study is to measure the $L_{\mathrm{X}}$-SFR-$Z$ relation at these redshifts, compare it to previous measurements at $z=0$ and $z=2$, and if the relation appears to be redshift-independent, use all the currently available data to better constrain the $Z$ dependence of HMXBs. In \S\ref{sec:data}, we describe the hCOSMOS and zCOSMOS spectroscopic surveys from which we selected our galaxy sample and the X-ray data from the \textit{Chandra} COSMOS Legacy survey we used in our analysis. \S\ref{sec:method} describes how we measured galaxy properties and calculated stacked X-ray luminosities. In \S\ref{sec:results}, we present our results on the $L_{\mathrm{X}}$-SFR-$Z$ relation across redshift. We summarize our findings in \S\ref{sec:conclusions} and briefly discuss the implications of these results for models of stellar evolution and studies which rely on estimates of HMXB emission.  Throughout this study, we assume a cosmology with $\Omega_m = 0.3$, $\Omega_{\Lambda} = 0.7$, and $h = 0.7$ and adopt the solar abundances from \citet{asplund09} ($Z_{\odot}=0.0142$, 12+log(O/H)$_{\odot}=8.69$).

\section{Data}
\label{sec:data}
Studying the relationship between $L_{\mathrm{X}}^{\mathrm{HMXB}}$, SFR, and $Z$ requires a large sample of star-forming galaxies with: 1) rest-frame optical spectra from which oxygen abundance can be measured, 2) sensitive X-ray data ($L_{\mathrm{X,lim}}\lesssim10^{42}$ erg s$^{-1}$) to screen out X-ray active galactic nuclei (AGN) and detect HMXB-dominated galaxies via X-ray stacking, and 3) sufficient photometric coverage to estimate SFRs through SED-fitting or spectroscopic coverage of emission line SFR indicators such as H$\alpha$.  The multi-wavelength coverage of the COSMOS field \citep{scoville07} offers suitable datasets for such studies.  We use two spectroscopic surveys (hCOSMOS and zCOSMOS), the \textit{Chandra} COSMOS Legacy survey, and the multiwavelength photometry available in the COSMOS field \citep{laigle16} to perform this study.  Here we briefly describe these data sets.  
\subsection{The hCOSMOS survey}
One of our galaxy samples is taken from the hCOSMOS survey, which was performed using the Hectospec multifiber spectrograph on the 6.5m MMT (\citealt{fabricant98}; \citealt{fabricant05}).  Hectospec covers the wavelength range $3700-9100$ {\AA} at a resolution of $R\sim1500$.  The hCOSMOS survey targeted galaxies in the COSMOS field with $r<21.3$, and obtained 4362 reliable spectroscopic redshifts, including 1701 new redshifts \citep{damjanov18}.  Hectospec data were reduced with HSRED v2.0, developed by the SAO Telescope Data Center.  Redshifts were measured by cross-correlating the observed spectra against a library of template spectra \citep{kurtz98}.  The majority of the galaxies observed by Hectospec are at $0.1<z<0.4$.  The details of Hectospec spectral calibration and spectroscopic redshift determination are provided in \citet{damjanov18}.  

\subsection{The zCOSMOS survey}
Our second galaxy sample is selected from the zCOSMOS-bright survey, which obtained spectra of 20,000 galaxies with $i<22.5$ and $0.1<z<1.2$ \citep{lilly07}.  The zCOSMOS survey was performed with the VIMOS spectrograph \citep{lefevre03} on the 8m ESO VLT.  The VIMOS MR grism provides $R\sim600$ resolution over the spectral range $5550-9650${\AA}.  The calibration of zCOSMOS spectra and measurement of spectroscopic redshifts is described in detail in \citep{lilly07}.  In this study, we use the subsample of 939 zCOSMOS galaxies at $0.5<z<0.9$ for which \citet{maier15} measured metallicities.  Metallicity measurements are also available for 164 zCOSMOS galaxies at $z<0.4$ \citep{cresci12}.  However, we did not use these galaxies in our analysis because, due to significant overlap between the zCOSMOS and hCOSMOS samples, including the unique zCOSMOS $z<0.4$ galaxies would only increase the sample size by 6\% while introducing heterogeneity in the measurement of galaxy properties.  

\subsection{The Chandra COSMOS Legacy survey}
Both of our spectroscopic galaxy samples reside in the COSMOS field, which covers roughly 2 deg$^2$ and was observed by the \textit{Chandra X-ray Observatory} to a fairly uniform depth of approximately 160~ks.  The \textit{Chandra} COSMOS Legacy survey detected 4016 X-ray sources, reaching limiting fluxes of $2.2\times10^{-16}$,$1.5\times10^{-15}$, and $8.9\times10^{-16}$ erg s$^{-1}$ cm$^{-2}$ in the $0.5-2$, $2-10$, and $0.5-10$ keV bands \citep{civano16}.  For an absorbed power-law spectrum with a photon index of $\Gamma=1.4$ (the photon index of the cosmic X-ray background) and Galactic obscuration of $N_{\mathrm{H}}=2.6\times10^{20}$ cm$^{-2}$, these sensitivity limits correspond to rest-frame $2-10$ keV limiting luminosities of $L_{\mathrm{X}}\sim10^{41}-10^{43}$ for $z=0.1-0.9$.  Thus, this survey is sufficiently deep as to be able to detect all moderate and bright luminosity X-ray AGN over the redshift range spanned by our galaxy samples, which is important for our analysis (see \S\ref{sec:agn} for details).

\subsection{Multiwavelength photometry}
To derive the $M_*$ and SFR of each galaxy, we make use of the most recent COSMOS photometric catalog from \citep{laigle16}.  This catalog includes the near-UV band from \textit{GALEX}, the $u*$ band from the Canada-France-Hawaii Telescope (CFHT/MegaCam), five Subaru Suprime-Cam bands($B$, $V$, $r$, $i$, $z+$), four UltraVista bands($Y$, $H$, $J$, $Ks$), and four \textit{Spitzer}/IRAC bands (3.6, 4.5, 5.8, and 8.0$\mu$m).  To constrain the MIR and FIR emission as much as possible, we also use the 24$\mu$m and 70$\mu$m Multiband Imaging Photometer for \textit{Spitzer} (MIPS) bands (\citealt{sanders07}; \citealt{lefloc09}), as well as the \textit{Herschel Space Observatory} PACS (100$\mu$m, 160$\mu$m) and SPIRE (250$\mu$m, 350$\mu$m, 500$\mu$m) bands (\citealt{griffin10}; \citealt{pilbratt10}; \citealt{poglitsch10}).

\subsection{Comparison samples}
\label{sec:comparison}
In \S\ref{sec:results}, we use measurements of $L_{\mathrm{X}}$/SFR from additional samples of star-forming galaxies at different redshifts to investigate the connection between  $L_{\mathrm{X}}$/SFR, $Z$, and redshift.  \citetalias{fornasini19} studied the $Z$ dependence of HMXBs in $z\sim2$ galaxies, and we use the $L_{\mathrm{X}}$/SFR values they measure as a function of O/H for high sSFR (sSFR$>10^{-8.8}$ yr$^{-1}$) galaxies.  \citetalias{fornasini19} measure SFRs from the H$\alpha$ luminosity and apply a $Z$-dependent conversion factor, assuming the \citet{hao11} conversion factor appropriate for $Z=0.02$ for galaxies with 12+log(O/H) $>8.3$ and adjusted for $Z=0.004$ using the conversion factor from \citet{reddy18} for galaxies with 12+log(O/H) $<8.3$.  Their $L_{\mathrm{X}}$/SFR results derived using H$\alpha$ SFRs are consistent within 0.15 dex with results based on SED-derived SFRs assuming $Z=0.02$ BC03 stellar models and the \citet{calzetti00} extinction curve. \par
We also include results from four different studies of $z=0$ galaxies as points of comparison.  All four of these studies are based on individually X-ray detected nearby galaxies and use UV+IR SFRs.  One of these studies is \citetalias{brorby16}, which combined new and archival samples of nearby galaxies with O/H measurements to measure the local $L_{\mathrm{X}}$-SFR-$Z$ relation.  Two of the studies, \citetalias{lehmer10} and \citetalias{mineo12}, did not include $Z$ information in measuring the scaling relation between $L_{\mathrm{X}}$ and SFR.  However, \citetalias{fornasini19} calculated the average O/H of the galaxies used in these studies by combining O/H measurements available in the literature and O/H estimates based on the $M_*$-$Z$ relation from \citet{kewley08}.  The fourth study, \citetalias{lehmer19}, performed subgalactic modeling of the HMXB and LMXB scaling relations in order to better disentangle their contributions compared to previous works.  This study provides an estimate of $L_{\mathrm{X}}^{\mathrm{HMXB}}$/SFR for their complete sample of 38 nearby galaxies as well as an estimate based on a ``cleaned'' subsample.  This subsample excludes galaxies with lower $Z$ or a higher frequency of globular clusters compared to the bulk of the sample, because such galaxies are found to be significant outliers to the scaling relations. We use this ``cleaned'' scaling relation as a point of comparison for our results.  We calculated the mean O/H of this sample using the O/H values based on strong line indicators compiled in \citetalias{lehmer19} and O/H estimates based on the $M_*$-$Z$ relation from \citet{kewley08} for galaxies lacking strong line measurements; these choices were made for consistency with the other mean O/H estimates calculated by \citetalias{fornasini19} for the other $z=0$ samples.  We ensured that all O/H values were properly converted to the \citet{pettini04} O3N2 metallicity scale using the prescriptions from \citet{kewley08}.  \par
We made some corrections to the $L_{\mathrm{X}}$/SFR measurements from $z=0$ studies to allow a more direct comparison to our results.  The SFR values from the local studies were converted to be consistent with a Chabrier IMF.  In addition, the \citetalias{mineo12}, \citetalias{brorby16}, and \citetalias{lehmer19} $L_{\mathrm{X}}$ values were converted from the $0.5-8$ to $2-10$ keV band.  To calculate the $L_{\mathrm{X}}$ conversion factors, we assumed an absorbed power-law spectrum and the average $\Gamma$ and $N_{\mathrm{H}}$ parameters used in these respective studies, which span the ranges of $\Gamma=1.7-2.0$ and $N_{\mathrm{H}}\leq3\times10^{21}$ cm$^{-2}$.

\section{Methods}
\label{sec:method}
\subsection{Measurement of galaxy properties}
\subsubsection{SED-derived $M_*$ and SFR}
\label{sec:sedsfr}
We derived $M_*$ and SFR for all the galaxies in the hCOSMOS and zCOSMOS samples in a consistent way by fitting their SEDs following the procedure described in \citet{suh17}, which is briefly summarized here.  Each near-UV to far-IR SED was decomposed into a galactic stellar population and a starburst component representing the infrared emission from dust heated by the UV light from young stars; unlike \citet{suh17}, we did not include an AGN component because we removed any confirmed AGN from our sample (as described in \S\ref{sec:agn}).  We generated a set of synthetic spectra to represent the stellar component from the population synthesis models of \citeauthor{bruzual03} (2003; hereafter BC03).  We assumed the \citet{chabrier03} initial mass function (IMF), $Z=0.02$, the \citet{calzetti00} dust extinction law, and considered both exponentially decaying and constant star formation histories (SFHs).  To represent the cold dust emission, we used starburst templates from \citet{chary01} and \citet{dale02}.  For each galaxy, the spectroscopic redshift from hCOSMOS or zCOSMOS was adopted as a fixed input to the SED-fitting.  The best-fit template combination was determined using $\chi^2$ minimization, while the most representative value for each physical parameter and corresponding uncertainties were determined via Bayesian estimation.  \par
The $M_*$ of each galaxy is calculated by integrating the SFH of the best-fitting template.  In order to account for both obscured and unobscured star formation, we do not simply use the SFRs derived from the SFHs, but combine the contributions from the UV and IR luminosities, using the following relation from \citet{arnouts13}, which is adjusted for a \citet{chabrier03} IMF:
\begin{equation}
    \mathrm{SFR}_{\mathrm{total}} (M_{\odot} \mathrm{yr}^{-1}) = 8.6\times10^{-11}\times(L_{\mathrm{IR}}/L_{\odot}+2.3\times \nu L_{2300})
\end{equation}
where $L_{\mathrm{IR}}$ is the rest-frame $8-1000\mu$m star-forming IR luminosity integrated from the starburst template, and $L_{2300}$ represents the rest-frame absorption-corrected near-UV luminosity at 2300 \AA\hspace{1pt} in units of $L_{\odot}$.  For galaxies which are undetected at MIR and FIR wavelengths, we calculate a lower bound to the SFR based on the UV luminosity alone, and an upper bound by combining the UV contribution with the upper limits in the MIR-FIR bands.  For these galaxies, we also adopt a best estimate SFR value derived by multiplying their UV luminosity by the median ratio of the total SFR to the UV-only SFR as a function of E(B-V) for the MIR-FIR detected galaxies, where E(B-V) is determined from the SED-fitting.  This median ratio varies from $0.07-0.31$.  Our results are not significantly changed if we ignore the variation of SFR$_{\mathrm{total}}$/SFR$_{\mathrm{UV}}$ with E(B-V) and instead adopt the same median ratio for all galaxies that are not detected at MIR-FIR wavelengths.  If this SFR estimate exceeds the upper bound set by the \textit{Spitzer} MIPS and \textit{Herschel} limits, then we adopt that upper bound as the best estimate SFR.  In \S\ref{sec:systematic} we investigate how these uncertainties in the SED-derived SFRs impact our results.  \par
\begin{figure}
    \centering
    \includegraphics[width=0.5\textwidth]{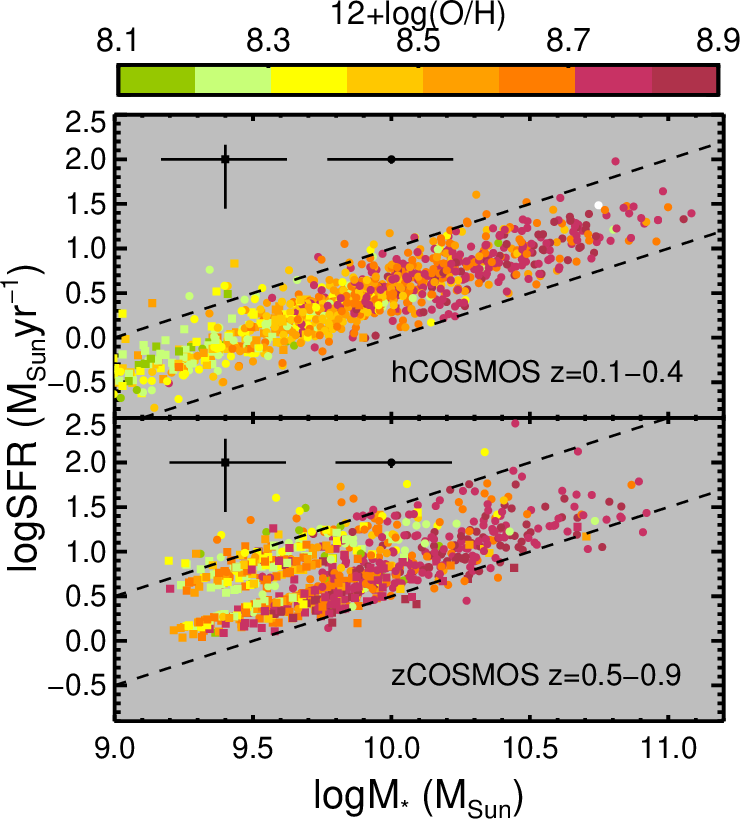}
    \caption{Best estimates of SFR versus $M_*$ for our galaxy samples taken from the hCOSMOS and zCOSMOS surveys are shown in the top and bottom panels, respectively.  Each galaxy is colored according to its metallicity.  Galaxies that are detected at MIR-FIR wavelengths are shown by circles, with their median statistical $M_*$ and SFR errors represented by the black circle with error bars.  Galaxies that are undetected at MIR-FIR wavelengths are shown by squares, with their median $M_*$ statistical error and the median difference between their upper and lower SFR bounds shown by the black square with error bars.  The dashed lines show the sSFR selection cuts we make as described in \S\ref{sec:agn}.}
    \label{fig:sfrmass}
\end{figure}

Figure \ref{fig:sfrmass} shows the SFR versus $M_*$ measurements for non-AGN galaxies from the hCOSMOS and zCOSMOS samples.  Galaxies that are detected at MIR-FIR wavelengths are shown by circles, while those that are undetected at MIR-FIR wavelengths are shown by squares.  We find that our $M_*$ and SFR values are in good agreement with measurements from other studies that use different SED fitting codes (\citealt{cresci12}; \citealt{laigle16}; \citealt{damjanov18}).

\subsubsection{Metallicity}
\label{sec:metallicity}
As done in previous studies of the $Z$ dependence of HMXBs (e.g. \citetalias{brorby16}; \citetalias{fornasini19}), we use the oxygen abundance (O/H) of star-forming H{\small II} regions as a proxy for the $Z$ of HMXBs, which are young stellar systems.  For both the hCOSMOS and the zCOSMOS samples, O/H measurements are derived using the $R_{23}$ line ratio, which is defined as:
\begin{equation}
    R_{23} = \frac{\mathrm{[O II]}\lambda3727+\mathrm{[OIII]}\lambda4959,5007}{\mathrm{H\beta}}
\end{equation}
\par
For the hCOSMOS galaxies, we fit the continuum around each of the emission lines required for the $R_{23}$ indicator using a linear polynomial.  We then normalize the spectral region containing the emission line to the fitted continuum, fit the line with a Gaussian profile, and measure the equivalent width (EW) of the line.  To calculate the $R_{23}$ line ratio, we use the EWs rather than fluxes of the lines because this method has been found to be reliable and less sensitive to interstellar reddening \citep{kobulnicky03}.  To ensure the O/H measurements are robust, we require S/N$>3$ in the [OII]$\lambda3727$ and H$\beta$ emission lines.  We correct for Balmer absorption by fitting stellar population synthesis models following \citet{zahid13}.  \par
To convert the $R_{23}$ ratio to an O/H measurement, we use the calibration of \citet{kk04}.  Metallicity is not a monotonic function of $R_{23}$, and the degeneracy between the upper and lower value solutions is usually broken by using another line ratio, typically [NII]/H$\alpha$.  However, since these lines are affected by red light leak or are redshifted out of the Hectospec spectral window for the bulk of the hCOSMOS sample, we assume the galaxies lie on the $R_{23}$ upper branch, which is likely a robust assumption on average for these low redshifts ($z=0.1-0.4$) and $M_*\gtrsim10^{9}M_{\odot}$ \citep{zahid11}.  We are able to estimate O/H for 1257 hCOSMOS galaxies.  Our O/H measurements reproduce the $M_*-Z$ relation measured by \citet{zahid13} at $z\sim0.29$.  For the subsample of 83 galaxies that are in common between the hCOSMOS and zCOSMOS surveys, our O/H values are consistent with measurements from \citet{cresci12} within the expected $\pm0.1-0.2$ dex uncertainties.  Only three of these galaxies exhibit O/H measurement differences significantly in excess of what would be expected due to statistical uncertainty; in all three cases, our O/H measurements are higher than those of \citet{cresci12} by $>0.4$ dex.  Since \citet{cresci12} base their O/H measurements on the [NII]/H$\alpha$ indicator, which does exhibit the same degeneracy as $R_{23}$, it is possible that these three galaxies should lie on the lower $R_{23}$ branch.  Given that this subsample is fairly representative of the full hCOSMOS sample in terms of $M_*$ which correlates with $Z$, the fact that only 4\% of the galaxies in this subsample show significant discrepancies with the \citet{cresci12} measurements suggests that our assumption that all galaxies lie on the upper $R_{23}$ branch does not introduce significant bias.  \par
For the zCOSMOS galaxies, we use the O/H measurements from \citet{maier15}. These O/H measurements are also based on the $R_{23}$ line ratio, but \citet{maier15} use the \citet{kd02} calibration.  For 39 of their galaxies, \citet{maier15} obtain near-IR spectra to break the degeneracy between O/H and $R_{23}$ using [NII]/H$\alpha$; based on this subsample of 39 galaxies, they find a trend between the Dn4000 index, [OIII]$\lambda$5007/H$\beta$, and whether galaxies lie on the upper/lower $R_{23}$ branch which is then used to break the $R_{23}$ degeneracy for the remaining 900 galaxies in their sample.  About 20\% of the zCOSMOS galaxies in \citet{maier15} are found to lie on the lower $R_{23}$ branch based on these observed trends.  Since these trends are likely to vary with redshift \citep{maier15}, we cannot apply the same criteria to our hCOSMOS sample at $z=0.1-0.4$.  Nonetheless, we note that, considering the redshift evolution of the $M_*-Z$ relation \citep{zahid13}, it is likely that the fraction of hCOSMOS galaxies falling on the lower $R_{23}$ branch is lower than for the zCOSMOS sample. \par
In order to compare the results based on the hCOSMOS and zCOSMOS samples to one another and to previous studies of the $L_{\mathrm{X}}$-SFR-$Z$ relation, we convert all O/H measurements to the \citet{pettini04} calibration for the O3N2 indicator using the conversions established by \citet{kewley08}.

\subsection{Sample Selection}
\label{sec:agn}
In order to study the X-ray emission from HMXBs, it is important to remove contamination from AGN as much as possible. Therefore, we identified AGN candidates through multiwavelength diagnostics and removed them from our galaxy samples.  We identified any individually detected X-ray sources with $L_{\mathrm{X}}>10^{42}$ erg s$^{-1}$ from \citet{civano16} as X-ray AGN.  We further excluded any X-ray sources with optical counterparts that were identified as optical AGN based on the photometric or spectroscopic classification of \citet{marchesi16}.  IR AGN were identified based on the criteria of \citet{donley12} and removed from the sample.  Finally, we also excluded optical AGN identified via optical line ratio diagnostic diagrams.  In the case of hCOSMOS sources, we identified optical AGN as sources lying above the \citet{kauffmann03} line in the [OIII]/H$\beta$ versus [NII]/H$\alpha$ diagram.  In the case of zCOSMOS, we excluded optical AGN using the blue diagnostic diagram based on [OII], H$\beta$, and [OIII] (\citealt{lamareille10}; \citealt{perez13}) as done by \citet{maier15}. \par
In addition to removing contamination from AGN, it is also important to limit contamination from low-mass X-ray binaries (LMXBs).  Whereas $L_{\mathrm{X}}^{\mathrm{HMXB}}$ is proportional to SFR, $L_{\mathrm{X}}^{\mathrm{LMXB}}$ is proportional to $M_*$ (\citealt{gilfanov04}; \citealt{lehmer10}).  Thus, the relative contribution of HMXBs and LMXBs to the integrated $L_{\mathrm{X}}$ of a galaxy depends on the specific SFR (sSFR=SFR/$M_*$).  The sSFR value at which galaxies transition from being LMXB to HXMB dominated is found to increase with redshift \citep{lehmer16}.  Thus, to limit the contribution of LMXBs, it is important to select galaxies above the transition sSFR value appropriate for a given redshift.  Therefore, we used a lower threshold of log(sSFR)$=-10.0$ for the hCOSMOS sample and log(sSFR)$=-9.5$ for the zCOSMOS sample.  Above this threshold value, the total $L_{\mathrm{X}}$/SFR of galaxies as a function of increasing sSFR is expected to gradually asymptote to the value appropriate for a pure HMXB population; $L_{\mathrm{X}}$/SFR drops by about 0.3 dex as sSFR increases by a factor of 10 from the threshold value \citep{lehmer10}.  Therefore, to minimize variations in the sSFR distributions of galaxies in different $Z$ bins, we also set an upper threshold of log(sSFR)$=-9.0$ for the hCOSMOS sample and log(sSFR)$=-8.5$ for the zCOSMOS sample.
\par
Finally, we limited the hCOSMOS sample to log$(M_*/M_{\odot})>9.0$, for which O/H values are reliably found to correspond to the upper branch of the $R_{23}$ ratio, as discussed in \S\ref{sec:metallicity}.  We limited the zCOSMOS sample to log$(M_*/M_{\odot})>9.2$ because the $M_*$ distribution of galaxies with $Z$ measurements from \citet{maier15} drops off steeply below this value.   After applying these selection criteria, our hCOSMOS sample consists of 858 galaxies, 319 of which are at $z=0.1-0.25$ and 539 of which are at $z=0.25-0.4$, and the zCOSMOS sample consists of 786 galaxies at $z=0.5-0.9$.

\subsection{X-ray stacking analysis}
The vast majority of the galaxies in our samples are too faint to be individually detected by \textit{Chandra}.  Therefore, to determine the mean X-ray luminosity for each set of galaxies grouped by $z$, O/H, or other physical properties, we first stack all the undetected galaxies and calculate their $\langle L_{\mathrm{X}} \rangle$.  Then, we compute a weighted average of the stacked $\langle L_{\mathrm{X}} \rangle$ and the $L_{\mathrm{X}}$ measurements of individually detected galaxies as detailed below.  \par
To perform our X-ray stacking analysis, we use the \textit{Chandra} stacking tool CSTACK v4.32 (\texttt{http://cstack.ucsd.edu/cstack/}).  For each galaxy position, CSTACK provides the net (background subtracted) count rate in the $0.5-2$ and $2-8$ keV bands.  CSTACK defines the aperture region for each object as a circle with radius equal to the 90\% encircled counts fraction (ECF) radius ($r_{90}$). The background region is defined as a $30^{\prime\prime}\times30^{\prime\prime}$ 
area centered on each object; this region excludes a $7^{\prime\prime}$-radius circle around the object as well as circular regions around any detected X-ray sources with radii which depend on the net X-ray source counts.  For each object, CSTACK only uses the \textit{Chandra} observations in which the object is located within 8$^{\prime}$ of the aim point, such that $r_{90}<7^{\prime\prime}$.  The reliability of the CSTACK code has been carefully vetted by previous studies (\citealt{mezcua16}; \citealt{fornasini18}).\par
As reported in \citet{fornasini18} (hereafter \citetalias{fornasini18}), the CSTACK source apertures are typically larger than twice the effective radii of star-forming galaxies with $M_*>10^{9}M_{\odot}$ and $z>0.1$. Thus, the X-ray photometric information derived is representative of the galaxies as a whole rather than just the nuclear component.  We find that $<10$\% of the galaxies in our sample have a neighboring galaxy of similar or lower magnitude within $r_{90}$.  We expect that these neighboring galaxies will only impact the stacked $L_{\mathrm{X}}$ by $<0.05$ dex based on the estimates presented in \citetalias{fornasini18}. \par
For each stack of individually X-ray undetected galaxies, we calculate the probability that the source could be generated by a noise fluctuation of the local background, and convert this to a Gaussian-equivalent detection significance.  The robustness of this stacking method was thoroughly investigated by \citetalias{fornasini18}.  For each stack, we also compute the exposure-weighted average net count rates, and convert the $0.5-2$ keV count rate to a mean X-ray luminosity based on the X-ray spectrum described in \S\ref{sec:xspectrum} and using the mean redshift of the stack to compute the $k$-correction.  We also calculate the exposure-weighted means of galaxy properties (i.e. $M_*$, SFR, sSFR, O/H, $z$). Full details of all these calculations are provided in \citetalias{fornasini18}.  The mean $L_{\mathrm{X}}$/SFR for each stack of X-ray undetected galaxies is calculated as $\langle L_{\mathrm{X}} \rangle/\langle$SFR$\rangle$.  \citetalias{fornasini19} estimate that $\langle L_{\mathrm{X}} \rangle/\langle$SFR$\rangle$ should approximate $\langle L_{\mathrm{X}}$/SFR$\rangle$ with 0.05 dex accuracy based on the 0.3-0.4 dex of scatter observed in the local $L_{\mathrm{X}}$-SFR relations (\citealt{mineo12}; \citealt{brorby16}).  The 1$\sigma$ statistical uncertainties on the stacked count rate are calculated using a bootstrapping technique; the galaxies in each stack are resampled 1,000 times while the number of galaxies is conserved, and the stacking analysis is repeated in order to obtain the statistical distribution of the mean net count rate.  The errors associated with mean galaxy properties are also calculated using the bootstrapping technique described in \citet{lehmer16}.  \par
Having measured the mean $L_{\mathrm{X}}$ and galaxy properties of stacks of X-ray undetected galaxies, we combine this information with the $L_{\mathrm{X}}$ and galaxy property measurements of individually X-ray detected galaxies.  To calculate the mean $L_{\mathrm{X}}$ and galaxy properties of both the X-ray detected and undetected galaxies in a stack, we perform a weighted average in which the weights are the number of galaxies represented by a given measurement.  In estimating $\langle L_{\mathrm{X}}$/SFR$\rangle$ for a stack, we use $\langle L_{\mathrm{X}} \rangle/\langle$SFR$\rangle$ for the individually X-ray undetected galaxies and the $L_{\mathrm{X}}$/SFR measurements of individually X-ray detected galaxies.  The galaxy properties and X-ray properties of each stack are provided in Table \ref{tab:prop} and \ref{tab:xray}, respectively.

\subsubsection{Constraints on the X-ray spectrum}
\label{sec:xspectrum}
Converting the $0.5-2$ keV net count rates to $2-10$ keV rest-frame luminosities requires a spectral model.  Since our stacks do not possess sufficient X-ray counts for meaningful spectral analysis, we use hardness ratios to roughly constrain the average X-ray spectrum of our galaxies.  For each stack of X-ray undetected galaxies, we use the Bayesian estimation code BEHR \citep{park06} to estimate the hardness ratio defined as $HR=(H-S)/(H+S)$, where $H$ and $S$ represent the net count rates in the $2-8$ and $0.5-2$ keV bands, respectively.  \par
\begin{figure}
    \centering
    \includegraphics[width=0.45\textwidth]{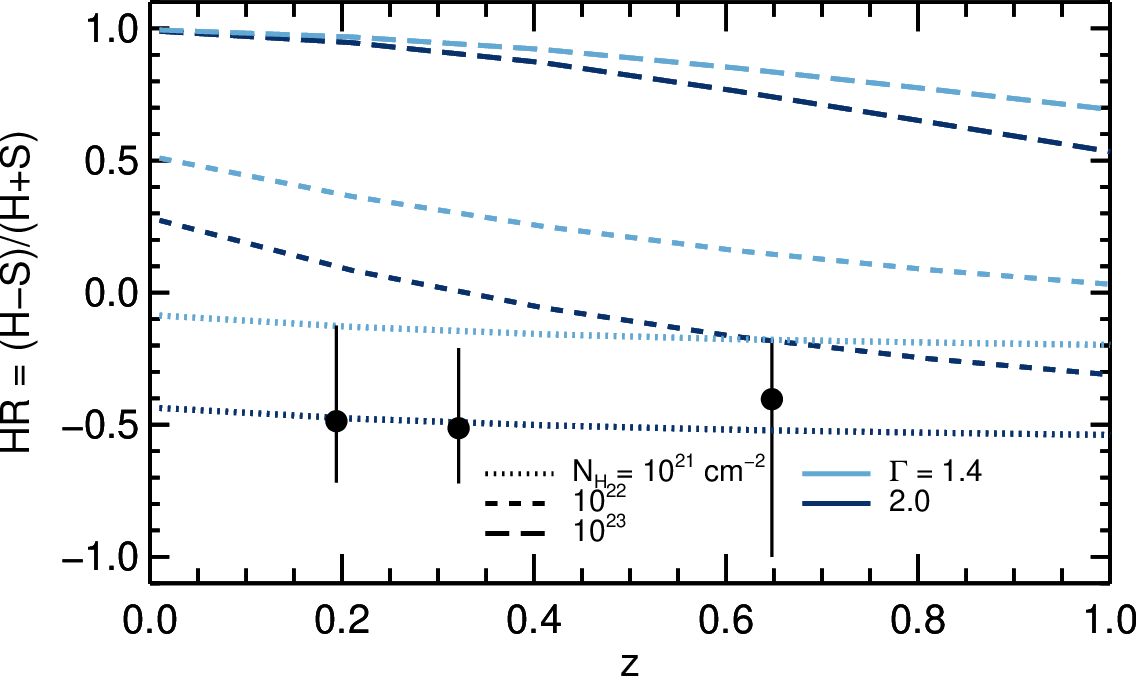}
    \caption{Hardness ratios versus redshift for stacks \#1-3, which split our galaxy samples by redshift alone and thus have the highest signal-to-noise ratio.  Hardness ratios are defined as $HR=(H-S)/(H+S)$ where $H$ and $S$ represent the net count rates in the $2-8$ and $0.5-2$ keV bands, respectively. Error bars represent 1$\sigma$ uncertainties.  Blue lines display the expected hardness ratios for sources with different absorbed power-law spectra; the line style represents different column densities ($N_{\mathrm{H}}$) while the line color represents different $\Gamma$. Our stacks are consistent with relatively unobscured ($N_{\mathrm{H}}<10^{22}$ cm$^{-2}$) spectra with $\Gamma=1.4-2.3$.}
\label{fig:hr}
\end{figure}

The HRs of stacks \#1-3 are shown in Figure \ref{fig:hr}.  The lines in Figure \ref{fig:hr} display the expected HRs for an absorbed power-law spectral model and different values of the column density ($N_{\mathrm{H}}$) and photon index ($\Gamma$).  The HRs of stacks \#1-3 are in good agreement with an aborbed power-law model with low obscuration ($N_{\mathrm{H}}\lesssim10^{22}$ cm$^{-2}$) and $\Gamma\approx1.4-2.3$, a typical range of photon indices for HMXBs.  The HRs of stacks \#4-13 and of individually X-ray detected sources are also statistically consistent with these values.
\par
For our default spectral model, we adopt $N_{\mathrm{H}}=10^{21}$ cm$^{-2}$ and $\Gamma=2$, which is the same model assumed by previous studies of the HMXB $Z$-dependence (\citetalias{brorby16}; \citetalias{fornasini19}). Varying $N_{\mathrm{H}}$ and $\Gamma$ within the ranges consistent with the data results in differences of $\pm0.3$ dex in $L_{\mathrm{X}}$.  Following \citetalias{fornasini18}, if we consider a more complex spectral model which includes a thermal \texttt{apec} component representing the hot gas in a galaxy and assume that the $L_{\mathrm{X}}$ contribution of this gas obeys the scaling relation with SFR parameterized by \citet{mineo12b}, the absorbed power-law model representing the HMXB population would be consistent with higher $N_{\mathrm{H}}\approx5\times10^{21}$ cm$^{-2}$ and $\Gamma\approx2$.  Adopting this more complex spectral model does not significantly change our estimate of the rest-frame $2-10$ keV luminosities.

\begin{table*}
\begin{minipage}{\textwidth}
\centering
\footnotesize
\caption{Mean Galaxy Properties of Stacks}
\begin{tabular}{ccccccccc} \hline \hline
\T \multirow{2}{*}{Stack ID} & \multicolumn{2}{c}{\# Galaxies} & \multirow{2}{*}{$z$ range} & \multirow{2}{*}{$\langle z \rangle$} & log$\langle M_* \rangle$ & \multirow{2}{*}{12+log(O/H)} & $\langle \mathrm{SFR}\rangle$ & log$\langle \mathrm{sSFR} \rangle$ \\  \cline{2-3}
& Detected & Stack & & & ($M_{\odot}$) & & ($M_{\odot}$ yr$^{-1}$) & (yr$^{-1}$) \\
\B (a) & (b) & (c) & (d) & (e) & (f) & (g) & (h) & (i) \\
\hline
\multicolumn{5}{l}{\textbf{Redshift binning}} \\
\hline
\T 1 & 4 & 393 & $0.1-0.25$ & 0.19 & $9.925^{+0.022}_{-0.007}$ & $8.512^{+0.003}_{-0.001}$ & $2.645^{+0.008}_{-0.010}$ & $-9.483^{+0.029}_{-0.008}$\\
\T 2 & 3 & 573 & $0.25-0.4$ & 0.32 & $10.213^{+0.024}_{-0.008}$ & $8.587^{+0.002}_{-0.001}$ & $5.845^{+0.017}_{-0.021}$ & $-9.405^{+0.028}_{-0.008}$\\
\T \B 3 & 0 & 786 & $0.5-0.9$ & 0.65 & $9.986^{+0.029}_{-0.008}$ & $8.615^{+0.003}_{-0.001}$ & $9.084^{+0.034}_{-0.035}$ & $-8.968^{+0.030}_{-0.007}$\\
\hline
\multicolumn{5}{l}{\textbf{Metallicity binning}} \\
\hline
\T 4 & 0 & 70 & $0.1-0.25$ & 0.18 & $9.393^{+0.030}_{-0.008}$ & $8.235\pm0.001$ & $0.780^{+0.004}_{-0.003}$ & $-9.481^{+0.031}_{-0.008}$\\
\T 5 & 2 & 127 & $0.1-0.25$ & 0.19 & $9.618^{+0.033}_{-0.008}$ & $8.459\pm0.002$ & $1.337^{+0.006}_{-0.005}$ & $-9.490^{+0.040}_{-0.009}$\\
\T 6 & 1 & 86 & $0.1-0.25$ & 0.20 & $9.941^{+0.027}_{-0.008}$ & $8.612^{+0.002}_{-0.001}$ & $2.994^{+0.010}_{-0.014}$ & $-9.466^{+0.036}_{-0.009}$ \\
\T 7 & 1 & 110 & $0.1-0.25$ & 0.20 & $10.258^{+0.018}_{-0.007}$ & $8.754^{+0.002}_{-0.001}$ & $5.439^{+0.014}_{-0.019}$ & $-9.491^{+0.029}_{-0.008}$ \\
\T 8 & 0 & 58 & $0.25-0.4$ & 0.31 & $10.022^{+0.024}_{-0.008}$ & $8.263\pm0.001$ & $3.494^{+0.024}_{-0.018}$ & $-9.344^{+0.028}_{-0.008}$\\
\T 9 & 1 & 151 & $0.25-0.4$ & 0.32 & $9.913^{+0.033}_{-0.009}$ & $8.467^{+0.002}_{-0.001}$ & $3.282\pm0.012$ & $-9.394^{+0.034}_{-0.009}$ \\
\T 10 & 0 & 143 & $0.25-0.4$ & 0.32 & $10.177^{+0.025}_{-0.008}$ & $8.612^{+0.002}_{-0.001}$ & $5.722^{+0.018}_{-0.023}$ & $-9.398^{+0.028}_{-0.008}$ \\
\T 11 & 2 & 200 & $0.25-0.4$ & 0.33 & $10.392^{+0.023}_{-0.008}$ & $8.754\pm0.002$ & $8.393^{+0.017}_{-0.027}$ & $-9.436^{+0.028}_{-0.008}$ \\
\T 12 & 0 & 480 & $0.5-0.9$ & 0.65 & $9.811^{+0.028}_{-0.008}$ & $8.505^{+0.004}_{-0.002}$ & $7.200^{+0.031}_{-0.028}$ & $-8.912^{+0.029}_{-0.007}$\\
\T \B 13 & 0 & 306 & $0.5-0.9$ & 0.65 & $10.172^{+0.029}_{-0.008}$ & $8.778\pm0.001$ & $12.104^{+0.039}_{-0.045}$ & $-9.078^{+0.032}_{-0.008}$ \\

\hline \hline
\multicolumn{9}{p{6.0in}}{\T Notes:

(g) Mean O/H based on $R_{23}$ indicator converted to \citet{pettini04} calibration.  

(h) SED SFR listed assumes $Z=0.02$ and Calzetti extinction curve. 
} \\
\end{tabular}
\label{tab:prop}
\end{minipage}
\end{table*}

\begin{table*}
\begin{minipage}{\textwidth}
\centering
\footnotesize
\caption{Stacked X-ray Properties}
\begin{tabular}{cccccc} \hline \hline
\T \multirow{2}{*}{Stack ID} & Stack exposure & Stack & Stack & $\langle L_{\mathrm{X}} \rangle$ & \multirow{2}{*}{log$\langle\frac{ L_{\mathrm{X}}}{ \mathrm{SFR}}\rangle$}\\
 & (Ms) & Significance & Net counts & ($10^{40}$ erg s$^{-1}$) & \\
 (a) & (b) & (c) & (d) & (e) & (f) \\
 \hline
\multicolumn{5}{l}{\textbf{Redshift binning}} \\
\hline
\T 1 & 42.3 & 10.5 & $277\pm27$ & $0.70^{+0.07}_{-0.09}$ & $39.44\pm0.04$ \\
\T2 & 63.2 & 11.6 & $368\pm33$ & $1.77^{+0.18}_{-0.21}$ & $39.49\pm0.04$ \\
\T \B 3 & 82.8 & 4.3 & $149\pm34$ & $2.66^{+0.70}_{-0.67}$ & $39.47\pm0.10$ \\
 \hline
\multicolumn{5}{l}{\textbf{Metallicity binning}} \\
\hline
\T 4 & 8.2 & 4.1 & $48\pm12$ & $0.49^{+0.15}_{-0.18}$ & $39.80\pm0.12$ \\
\T 5 & 13.9 & 4.4 & $67\pm15$ & $0.59^{+0.12}_{-0.14}$ & $39.64\pm0.08$ \\
\T 6 & 9.3 & 5.6 & $69\pm13$ & $0.79^{+0.17}_{-0.21}$ & $39.41\pm0.09$ \\
\T 7 & 10.8 & 7.1 & $92\pm14$ & $0.93^{+0.18}_{-0.20}$ & $39.26\pm0.08$ \\
\T 8 & 6.2 & 4.0 & $39^{+11}_{-10}$ & $1.68^{+0.55}_{-0.46}$ & $39.68\pm0.12$ \\
\T 9 & 17.0 & 4.6 & $75\pm16$ & $1.27^{+0.32}_{-0.35}$ & $39.58\pm0.10$ \\
\T 10 & 16.5 & 6.2 & $101\pm17$ & $1.74\pm0.38$ & $39.48\pm0.09$ \\
\T 11 & 23.6 & 7.7 & $153\pm21$ & $2.17^{+0.31}_{-0.33}$ & $39.45\pm0.06$ \\
\T 12 & 51.0 & 3.1 & $86\pm27$ & $2.50^{+0.91}_{-0.89}$ & $39.54\pm0.13$ \\
\T \B 13 & 31.8 & 3.0 & $63\pm21$ & $2.91^{+1.04}_{-1.11}$ & $39.38\pm0.13$ \\
\hline \hline
 \multicolumn{6}{p{4.5in}}{\T Notes:
 
(c) Detection significance expressed in Gaussian $\sigma$ based on the probability that the source could be generated by a noise fluctuation of the local background.
 
 (d) Net counts of stacked, individually undetected galaxies in the $0.5-2$ keV band.  Errors are based on Poisson statistics.
 
 (e) Mean \textbf{$2-10$ keV} X-ray luminosity, including both individually detected and undetected sources.  Errors are based on bootstrapping.
 
 (f) Mean \textbf{$2-10$ keV} $L_{\mathrm{X}}$/SFR, including individually detected and undetected sources.
 }
 \\
\end{tabular}
\label{tab:xray}
\end{minipage}
\end{table*}

\begin{figure}
\centering
\includegraphics[width=0.5\textwidth]{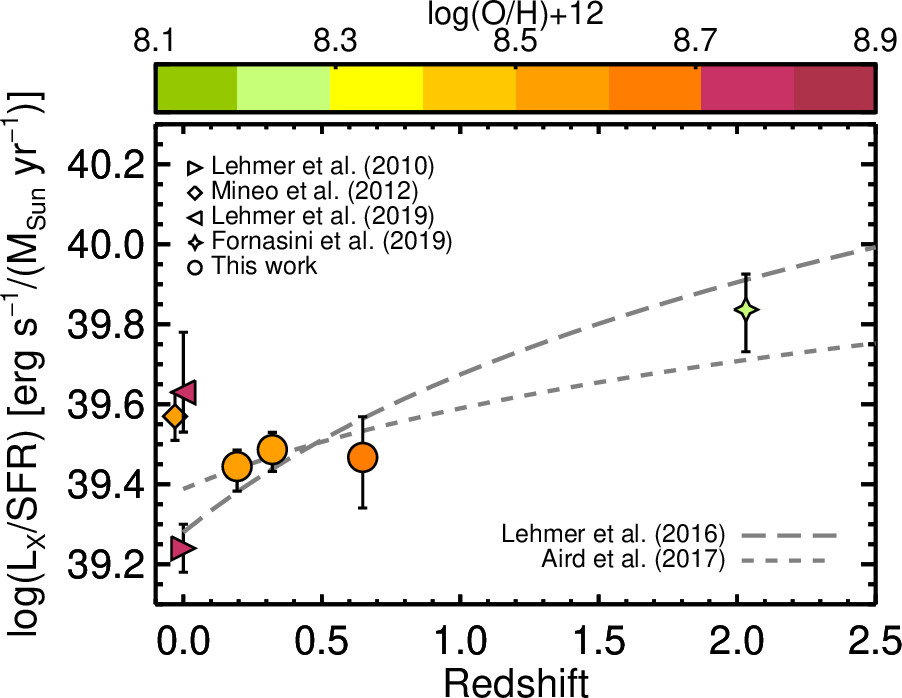}
\caption{Stacked $2-10$ keV $L_{\mathrm{X}}$/SFR values versus redshift of the hCOSMOS and zCOSMOS samples are shown by circles.  The hCOSMOS sample has been split into two redshift bins: $z=0.1-0.25$ and $z=0.25-0.4$.  Stars represent the stacked $L_{\mathrm{X}}$/SFR of high sSFR galaxies from \citetalias{fornasini19}.  The diamond and triangles represent local ($z=0$) measurements of the $L_{\mathrm{X}}$-SFR relation.  The symbol colors represent the mean O/H of the galaxy samples.  The long and short dashed lines display the redshift evolution of $L_{\mathrm{X}}$/SFR for HMXBs measured by \citet{lehmer16} and \citet{aird17}, respectively; since  the \citetalias{aird17} HMXB-only evolution is parametrized as a non-linear relation between $L_{\mathrm{X}}$ and SFR, this curve has been normalized for SFR$=6 M_{\odot}$ yr$^{-1}$, the mean SFR of our hCOSMOS and zCOSMOS galaxy samples.}
\label{fig:lxzevol}
\end{figure}

\begin{figure*}
\centering
\includegraphics[width=0.85\textwidth]{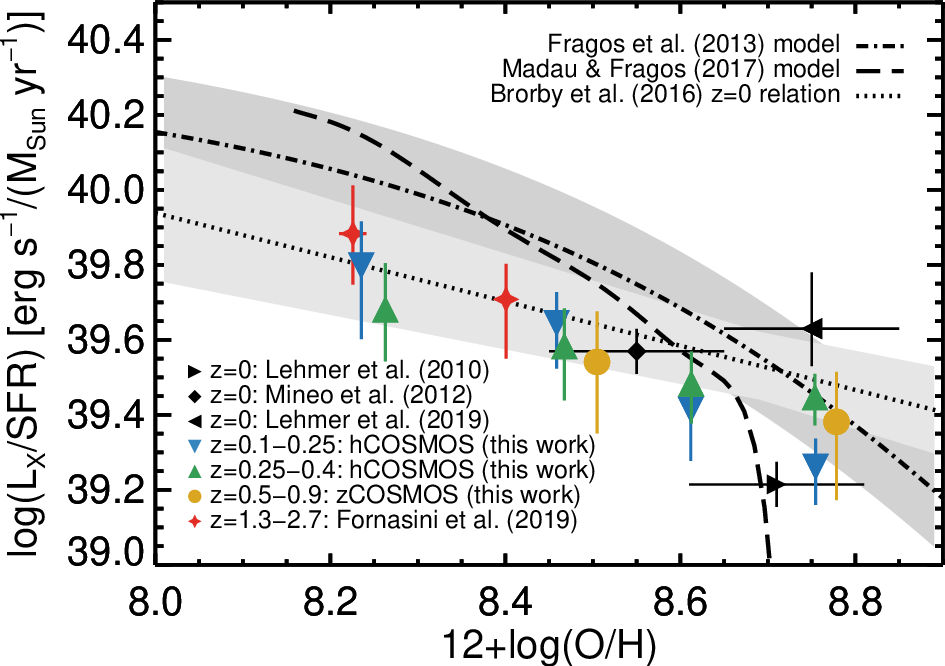}
\caption{Average $2-10$ keV $L_{\mathrm{X}}$/SFR versus O/H for galaxy samples in different redshift ranges are shown by different symbols/colors.  For the $z=0$ data points, since many of the O/H estimates are based on the $M_*-Z$ relation (see \ref{sec:comparison}), the horizontal error bars represent the scatter of $M_*-Z$ relation from \citet{kewley08}.}  The local $L_{\mathrm{X}}$-SFR-$Z$ relation from \citetalias{brorby16} is shown by the dotted line with corresponding error shown by the light gray shaded region.  The dash-dotted line represents the mean of the six highest likelihood models from \citetalias{fragos13b}, with the parameter space covered by these six models shown by the dark gray shaded region.  The dashed line shows the best-fit model from \citetalias{madau17}, which has been converted from the \citet{kk04} $R_{23}$ scale to the O3N2 scale from \citet{pettini04} using the prescription of \citet{kewley08}. 
\label{fig:lxzdep}
\end{figure*}

\begin{figure}
\centering
\includegraphics[width=0.45\textwidth]{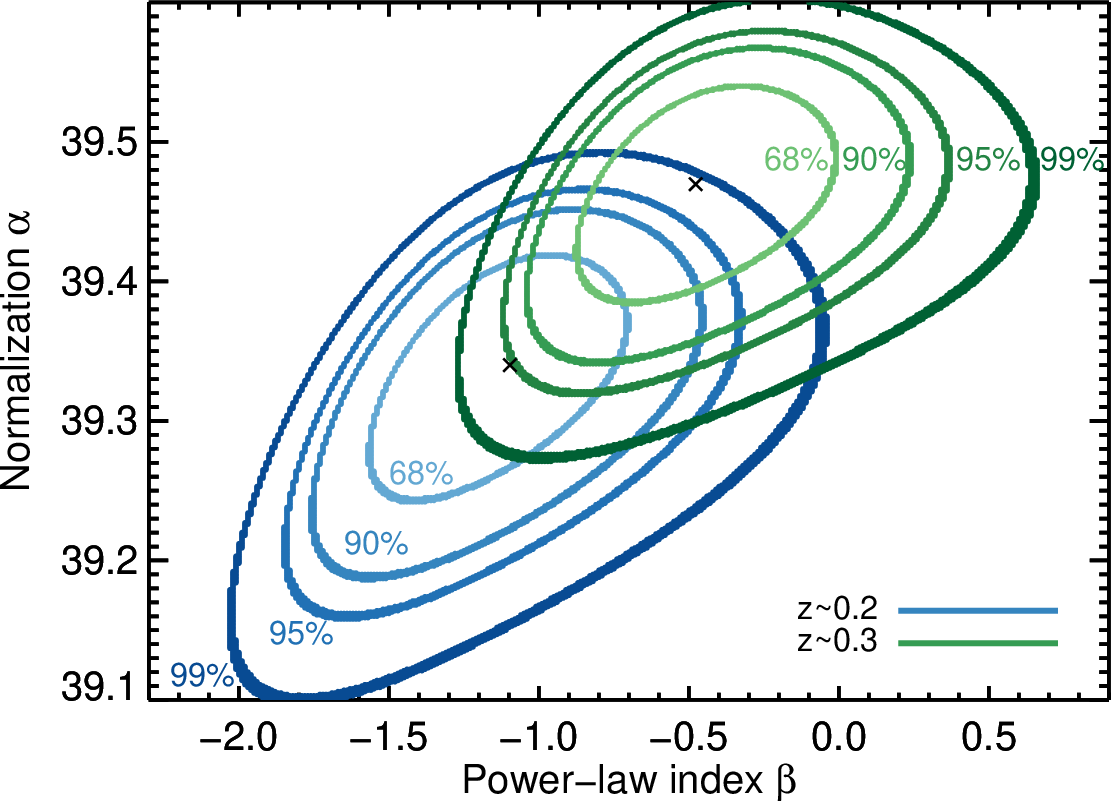}
\caption{Parameter confidence contours for the normalization (y-axis) and power-law index (x-axis) of the $L_{\mathrm{X}}$-SFR-$Z$ relation from Equation \ref{eq:sfrfixed} fit to the $z\sim0.2$ and $z\sim0.3$ stacks are shown in blue and green, respectively.  The best-fit values are indicated by cross symbols, and the contours shown represent 68\%, 90\%, 95\%, and 99\% confidence levels.}
\label{fig:confidence}
\end{figure}

\begin{figure*}
\centering
\includegraphics[width=0.85\textwidth]{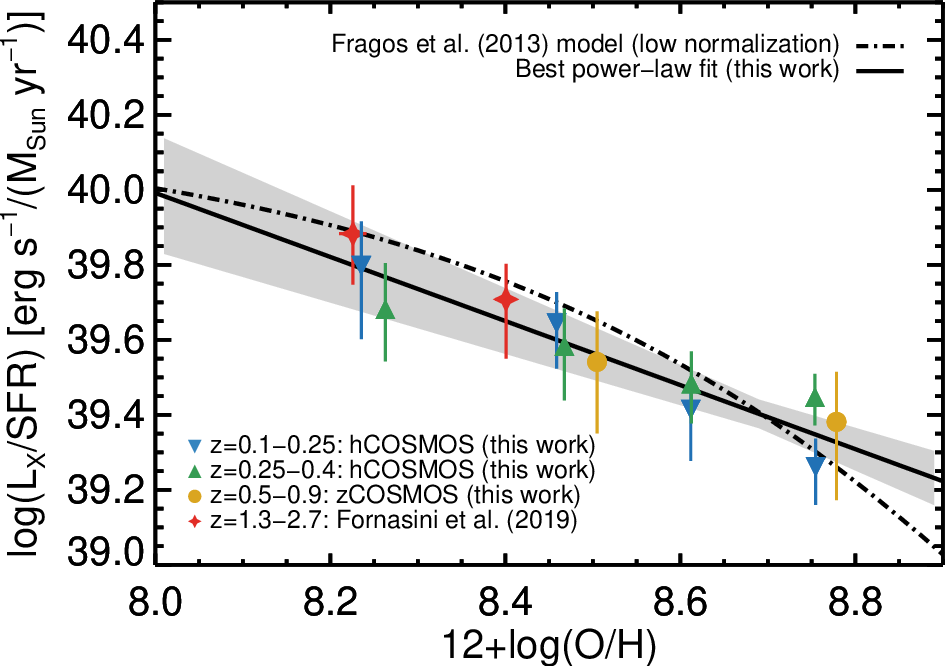}
\caption{Average $2-10$ keV $L_{\mathrm{X}}$/SFR versus O/H for galaxy samples in different redshift ranges are shown by different symbols/colors. Our best-fit power-law $L_{\mathrm{X}}$-SFR-$Z$ relation for the $z>0$ stacks is shown by the solid line, with corresponding error shown by the gray shaded region.  The lower bound of the \citetalias{fragos13b} models (which is 0.15 dex lower than the mean of their best-fitting models) also provides a good fit to the $z>0$ stacks and is shown by a dash-dotted line.}
\label{fig:bestfit}
\end{figure*}

\begin{figure*}
\centering
\includegraphics[width=0.85\textwidth]{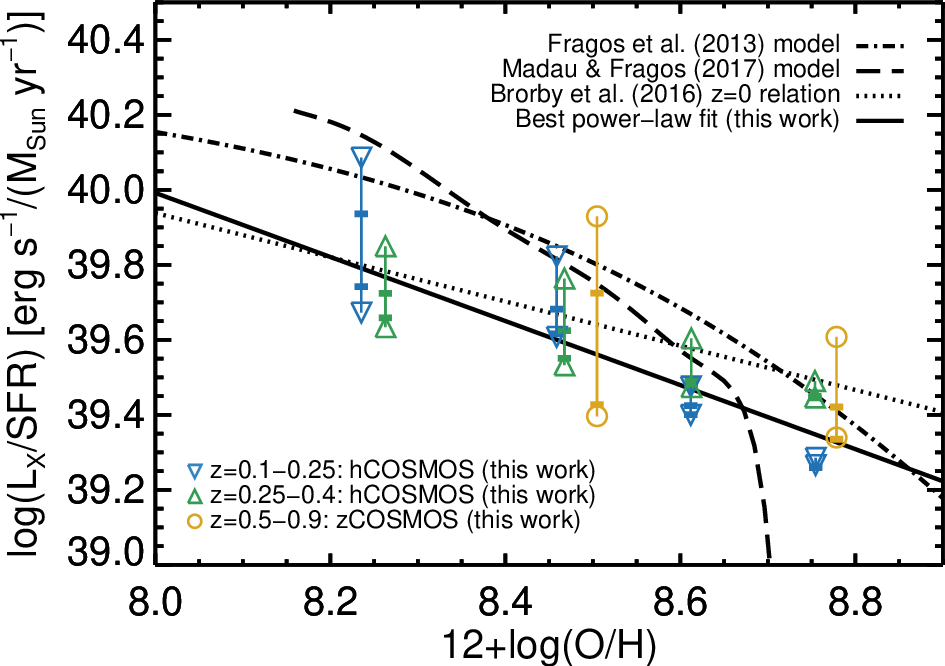}
\caption{Average $2-10$ keV $L_{\mathrm{X}}$/SFR upper/lower limits versus O/H for galaxy stacks from the hCOSMOS and zCOSMOS surveys in different redshift ranges are shown by different symbols/colors.  For each galaxy stack, the horizontal bar symbols show the upper/lower limits on $L_{\mathrm{X}}$/SFR based on adopting the UV-only lower bound on SFR or the \textit{Herschel}-limit upper bound on SFR, respectively, for galaxies which have no detections at wavelengths $\geq24\mu$m.  The open symbols show the upper/lower limits on $L_{\mathrm{X}}$/SFR based on adopting the lower/upper bounds on SFR for all galaxies which are undetected at wavelengths $\geq100\mu$m.  The lines are as described in Figures \ref{fig:lxzdep} and \ref{fig:bestfit}.}
\label{fig:range}
\end{figure*}

\section{Results and Discussion}
\label{sec:results}

\subsection{$L_{\mathrm{X}}$/SFR as a function of redshift}
We first checked whether our galaxies exhibit similar $L_{\mathrm{X}}$/SFR as a function of redshift as found by previous studies.  We divided our galaxies into three redshift bins ($z=0.1-0.25$, $0.25-0.4$, $0.5-0.9$), and stacked the X-ray data.  The properties of these stacks are listed in rows \#$1-3$ in Tables \ref{tab:prop} and \ref{tab:xray}, and their $L_{\mathrm{X}}$/SFR values versus redshift are shown in Figure \ref{fig:lxzevol} with circles. \par
 The $L_{\mathrm{X}}$/SFR of the $z\sim2$ sample from \citetalias{fornasini19} is significantly enhanced relative to our stacks at $z=0.1-0.9$.  Since $Z$ information is available for both these samples, together they can be used to test the connection between the redshift evolution and $Z$ dependence of HMXBs, which is discussed in \S\ref{sec:diffz}.  Our stacks are consistent with the $L_{\mathrm{X}}$/SFR values expected based on the redshift evolution relations of \citetalias{lehmer16} and \citetalias{aird17} shown by lines in Figure \ref{fig:lxzevol}, which are derived using galaxies at $z=0.1-4$.  The stack luminosities fall below the upper limit on XRB emission estimated using the redshift-dependent parametrization of \citetalias{fornasini18}.  We note that the hardness ratios of our stacks are lower than those of most of the stacks of star-forming galaxies from \citetalias{fornasini18} at similar redshift; as a key difference between our sample selection and that of \citetalias{fornasini18} is our rejection of sources with AGN signatures, it is possible that the higher hardness ratios of the \citetalias{fornasini18} sample are due to the presence of obscured AGN.   If the obscuration of the \citetalias{fornasini18} stacks is associated with AGN and not the HMXB populations, then the HMXB redshift evolution of $L_{\mathrm{X}}$/SFR measured by \citetalias{fornasini18} would be consistent with \citetalias{lehmer16} and \citetalias{aird17}.  However, since there are additional differences between the \citetalias{fornasini18} sample and ours (including our S/N requirements for emission lines which could bias us against galaxies with more obscured star-forming regions), we cannot definitively assess that the difference in the hardness ratios of the samples is due to the presence or absence of obscured AGN. \par
It is worth noting that our stacks and previous measurements from \citetalias{lehmer16} and \citetalias{aird17} at similar redshift lie in between the local ($z=0$) measurements of $\langle L_{\mathrm{X}}/\mathrm{SFR} \rangle$.  The fact that these $z\sim0.2-0.7$ measurements lie below some $z=0$ measurements does not necessarily rule out redshift evolution of $L_{\mathrm{X}}$/SFR.  The differences of $0.15-0.2$ dex between the $z=0$ measurements and our stacks could be accounted for by systematic effects, which are discussed in \S\ref{sec:systematic}.  In \S\ref{sec:diffz}, we discuss some possible explanations for the differences between the local measurements. \par

\subsection{The $Z$ dependence of $L_{\mathrm{X}}$/SFR at different redshifts}
\label{sec:diffz}
In order to study the relationship between $L_{\mathrm{X}}$/SFR and $Z$ in more detail, we divide the galaxies in each of our three redshift intervals by $Z$.  The properties of the resulting X-ray stacks are reported in rows \#4-13 of Tables \ref{tab:prop} and \ref{tab:xray}.  Figure \ref{fig:lxzdep} displays the $L_{\mathrm{X}}$/SFR versus $Z$ of these stacks, along with local measurements from \citetalias{mineo12}, \citetalias{lehmer10}, and \citetalias{lehmer19} and stacked measurements of $z\sim2$ galaxies from \citetalias{fornasini19}; the different colors and symbol shapes represent the different redshift intervals of the galaxy samples.  \par
Overall, these data from $z=0.1-2$ combined together show that a clear anti-correlation between $L_{\mathrm{X}}$/SFR and $Z$ exists and this relationship appears to be very similar across redshift.  The probability that $L_{\mathrm{X}}$/SFR and $Z$ are anti-correlated for the $z>0$ samples combined together is $>99.9$\% using both the Pearson correlation and Spearman rank correlation tests.  The strong agreement between $L_{\mathrm{X}}$/SFR values at similar O/H in galaxy samples spanning $z\approx0.2-2$ provides the strongest evidence to date that the redshift evolution of $L_{\mathrm{X}}$/SFR can be attributed to the $Z$ dependence of HMXB populations. 
\par
However, some $>1\sigma$ differences can be observed between some of the $L_{\mathrm{X}}$/SFR measurements at similar $Z$.  Therefore, it is important to test whether the relationship between $L_{\mathrm{X}}$/SFR and $Z$ exhibits any significant variation with redshift. \par
One potentially significant disagreement between the different redshift samples visible in Figure \ref{fig:lxzdep} is that the $z\approx0.2$ stacks exhibit a steeper anti-correlation between $L_{\mathrm{X}}$/SFR and $Z$ than the $z\approx0.3$ stacks.  There are sufficient stacks that we can independently fit the $L_{\mathrm{X}}$-SFR-$Z$ relation at these redshifts.  We find the best-fitting relation for these two sets of stacks using $\chi^2$ minimization and adopting a simple power-law relation of the form:
\begin{equation}
\frac{L_{\mathrm{X}}}{\mathrm{SFR}} = \alpha\left(\frac{(\mathrm{O/H})}{(\mathrm{O/H})_{\odot}}\right)^{\beta}
\label{eq:sfrfixed}
\end{equation}
For $z\approx0.2$, the best-fit values are log($\alpha$)$=39.34\pm0.06$ and $\beta=-1.10\pm0.29$, while for $z\approx0.3$, they are log($\alpha$)$=39.47\pm0.05$ and $\beta=-0.48\pm0.28$.   Thus, the $z\approx0.2$ stacks do favor a steeper $L_{\mathrm{X}}$-SFR-$Z$ relation, albeit weakly. As shown in Figure \ref{fig:confidence}, the 1$\sigma$ confidence contours of the fit parameters for the two samples overlap slightly, their 2$\sigma$ confidence contours overlap substantially, but the best-fit values are only consistent at $95-99$\% confidence.  Fitting both sets of stacks together results in a good fit with a reduced $\chi^2$ of 0.64 and best-fitting values of log($\alpha$)$=39.40\pm0.04$ and $\beta=-0.80\pm0.20$.  Thus, while the $z\approx0.2$ stacks favor a steeper relation than the $z\approx0.3$ stacks, this result is not very statistically significant. \par
Even though the evidence for a steeper $L_{\mathrm{X}}$-SFR-$Z$ relation at $z\approx0.2$ compared to $z\approx0.3$ is weak, it is worth considering whether a flattening of this relation with redshift could be explained by observational biases or is likely to be intrinsic. As discussed in \S\ref{sec:agn}, $L_{\mathrm{X}}$/SFR is known to vary with sSFR because the relative contributions of HMXBs and LMXBs change with sSFR.  Therefore, if the way in which the sSFR distributions vary with $Z$ differs at different redshifts, the $L_{\mathrm{X}}$-SFR-$Z$ relation could appear flatter or steeper.  However, we found that the difference in the $L_{\mathrm{X}}$/SFR-$Z$ slope for galaxy stacks below and above $z=0.25$ persisted regardless of whether we selected low sSFR or high sSFR galaxies or if we narrowed the sSFR range of our sample.  \citetalias{fornasini19} found that another factor which can affect the steepness of the $L_{\mathrm{X}}$-SFR-$Z$ relation is AGN contamination.  Like \citetalias{fornasini19}, we find that including known AGN in our stacks can change the stacked $L_{\mathrm{X}}$/SFR by $\pm0.1-0.2$ dex and overall flatten the anti-correlation between $L_{\mathrm{X}}$/SFR and $Z$.  At lower redshift, we can identify lower-luminosity AGN because most of the survey data we use for AGN identification are flux-limited.  Therefore, it would be natural to expect more AGN contamination in our higher redshift stacks, and this increase in AGN contamination at higher redshift could help explain the flatter $L_{\mathrm{X}}$/SFR-$Z$ slope observed at $z\approx0.3$ compared to $z\approx0.2$.  \par
Comparing the $z>0$ results to $z=0$ measurements, we find general agreement with the $L_{\mathrm{X}}$-SFR-$Z$ relation from \citetalias{brorby16}, with 10 of the 12 stacked values being within 1$\sigma$ of this relation.  However, all our stacks with 12+log(O/H)$\geq8.6$ fall below the \citetalias{brorby16} relation, and we calculate that the probability that the $z>0$ stacks are drawn from the \citetalias{brorby16} relation is 2.2\% assuming $\chi^2$ statistics.  The \citetalias{mineo12} $z=0$ $L_{\mathrm{X}}$/SFR measurement is in good agreement with our stacks with similar mean O/H.  At high O/H, there is some disagreement between different $z=0$ measurements of $L_{\mathrm{X}}$/SFR, and our $z>0$ stacks fall in between these measurements.  The \citetalias{lehmer10} value is consistent with the $z\approx0.2$ and $z\approx0.7$ stacks, while the \citetalias{lehmer19} value is higher than all our stacked values by $>90$\% confidence.  The discrepancy among these $z=0$ measurements and between them and the $z>0$ stacks at high O/H could result from several causes including: (i) the high obscuration of the LIRGs and ULIRGs in the \citetalias{lehmer10} sample, (ii) the average O/H of the $z=0$ samples not being properly estimated since for many of the galaxies in these samples we rely on the $M_*-Z$ relation and not on actual O/H measurements, (iii) the uncertainties on the average X-ray spectrum of our stacks and the $z=0$ galaxy samples, (iv) the heterogeneity of SFR calibrations used in different studies, and (v) selection effects that result in local samples being incomplete or unrepresentative.  These discrepancies highlight the need for further studies of the HMXB scaling relation at $z=0$ as a function of $Z$ with more complete samples.  Nonetheless, given all the sources of uncertainty listed above, we do not find any compelling evidence of variation of the $L_{\mathrm{X}}$-SFR-$Z$ relation between $z=0$ and the higher redshift samples.
\par
In summary, we find no strong evidence that the $L_{\mathrm{X}}$-SFR-$Z$ relation varies with redshift, and the weak trend that is seen between $z\approx0.2$ and $z\approx0.3$ could be explained by AGN contamination rather than being of intrinsic origin.  Thus, these combined data sets provide strong direct evidence that the observed redshift evolution of $L_{\mathrm{X}}$/SFR is driven by the $Z$ dependence of HMXBs. 

\subsection{Constraining the $L_{\mathrm{X}}$-SFR-$Z$ relation}
Having found no definitive evidence of redshift variation of the $L_{\mathrm{X}}$-SFR-$Z$ relation between $z\approx0.2$ and $z\approx2$, we proceed to jointly fit the galaxy samples across redshift to best constrain this relationship.  Using $\chi^2$ minimization, we first find the best relation for all $z>0$ stacks using the following functional form used by \citetalias{brorby16}:
\begin{equation}
L_{\mathrm{X}} = \alpha\left(\frac{(\mathrm{O/H})}{(\mathrm{O/H})_{\odot}}\right)^{\beta}\left(\frac{\mathrm{SFR}}{M_{\odot} \mathrm{yr}^{-1}}\right)^{\delta}
\end{equation}
The $z=0$ measurements are not included in this fit because: (i) the average O/H of the local galaxy samples are based on a mixture of actual O/H measurements and O/H estimates derived from the $M_*$-$Z$ relation and thus are more uncertain, and (ii) the local samples are subject to more complicated selection effects than our $z>0$ stacks.  The best-fit parameters are log($\alpha$)$=39.35\pm0.07$, $\beta=-0.91\pm0.17$, and $\delta=1.06\pm0.08$, resulting in a reduced $\chi^2=0.44$.  Since $\delta$ is consistent with a linear dependence of $L_{\mathrm{X}}$ on SFR, we fix it to 1 and fit the stacks using the functional form in Equation \ref{eq:sfrfixed}.  In this case, the best-fit parameters are log($\alpha$)$=39.40\pm0.04$, $\beta=-0.85\pm0.17$, and the reduced $\chi^2=0.46$.  This best-fit relation is shown in Figure \ref{fig:bestfit}.  \par
The $L_{\mathrm{X}}$/SFR-$Z$ slope ($\beta$) for the $z>0$ stacks is steeper than the slope found at $z=0$ by \citetalias{brorby16} ($\beta=-0.59\pm0.13$) but consistent with the slope measured by \citet{douna15} ($\beta=-1.01$) using a smaller sample of $z=0$ galaxies.  On the other hand, the normalization ($\alpha$) for the $z>0$ stacks is consistent with that of the \citetalias{brorby16} $z=0$ sample but higher than that found by \citet{douna15}.  Thus, the best-fit relation for the $z>0$ stacks at least falls within the range of $z=0$ measurements, although a more quantitative comparison of the agreement will require better understanding and controlling for the selection effects affecting local samples. \par
While a power-law relation between $L_{\mathrm{X}}$/SFR and $Z$ provides a good fit to the $z>0$ stacks, such a model is unphysical and likely would not extend to lower $Z$ than is probed by our stacks (12+log(O/H)$<8.0$).  Population synthesis models predict that the dependence of $L_{\mathrm{X}}$/SFR on $Z$ should flatten at low $Z$ because the wind mass loss rates become so low that the effects of $Z$ on stellar evolution saturate \citep{madau17}.  \citet{basu16} showed how variation of the normalization or slope of the HMXB luminosity function with $Z$ could reproduce the $Z$-dependence of $L_{\mathrm{X}}$/SFR predicted by theoretical models. \par
In order to provide more realistic constraints on the $L_{\mathrm{X}}$-SFR-$Z$ relation, we assess to what extent these theoretical models are consistent with our data.  We calculate the probability that the $z>0$ stacks are consistent with a given model assuming $\chi^2$ statistics.  The highest likelihood model from \citetalias{madau17} and the mean of the six highest likelihood models from \citetalias{fragos13b} are inconsistent with our stacks with $>99.9$\% probability.  However, we find that the lower bound of parameter space occupied by the \citetalias{fragos13b} models, which is about 0.15 dex lower in $L_{\mathrm{X}}$/SFR than the mean \citetalias{fragos13b} model, is in good agreement with the $z>0$ stacks.  The null hypothesis that the $z>0$ stacked data is consistent with this lower \citetalias{fragos13b} model is rejected only with 56\% probability, which is not significant.  This model is shown by a dash-dotted line in Figure \ref{fig:bestfit}.  \par
An important consequence of a redshift-independent $L_{\mathrm{X}}$-SFR-$Z$ relation is that the redshift evolution of $L_{\mathrm{X}}$/SFR measured by a given study will depend on the $Z$ distribution of the galaxy sample used, which is correlated with the $M_*$ and SFR distribution of the sample (e.g. \citealt{mannucci10}; \citealt{maier14}; \citealt{maier15}; \citealt{sanders15}).  Thus, some of the differences between previous measurements of the redshift evolution of $L_{\mathrm{X}}$/SFR (shown in Figure \ref{fig:lxzevol}) may be due to selection effects impacting the $Z$ distribution of the galaxy samples.

\subsection{Systematic effects}
\label{sec:systematic}
In comparing the data to population synthesis models, it is important to acknowledge the possible impact of systematic effects.  As discussed in \S\ref{sec:diffz}, we find that AGN contamination can have the effect of flattening the $L_{\mathrm{X}}$-SFR-$Z$ relation, as was found by \citetalias{fornasini19} for their $z\sim2$ sample.  It is therefore possible that the intrinsic relation is steeper than we observe.  \par
It is also important to consider the impact of uncertainties associated with SFR indicators on the observed slope and normalization of the $L_{\mathrm{X}}$-SFR-$Z$ relation.  One source of uncertainty is the fact that the total SED-derived SFRs of galaxies that are not detected at MIR-FIR wavelengths are not well constrained.  Strong constraints on the dust emission associated with obscured star formation requires detections at wavelengths $\geq100\mu$m, although data at 24$\mu$m and 70$\mu$m provides some constraining power on this emission.  In our hCOSMOS and zCOSMOS samples, 66\% of the galaxies are detected in at least one band $\geq24\mu$m, while only 32\% are detected at $\geq100\mu$m.  To assess the impact of uncertainties of the total SED-derived SFR on our results, we calculate upper and lower bounds on the average $L_{\mathrm{X}}$/SFR of each stack accounting for uncertainties in the SFR measurements of galaxies that are not detected at MIR-FIR wavelengths. We perform these calculations for two scenarios: a more liberal case in which we consider having at least one detection at $\geq24\mu$m as sufficient for constraining the IR dust emission, and a more conservative case in which we require at least one detection at $\geq100\mu$m to consider a galaxy's IR dust emission as well-constrained.  In calculating these upper and lower bounds on $L_{\mathrm{X}}$/SFR, we use the best estimate SFR values for galaxies considered to have well-constrained IR dust emission, and the lower/upper bounds on SFR described in \S\ref{sec:sedsfr} for galaxies which are considered to have poorly constrained IR dust emission. \par
The upper and lower bounds on $L_{\mathrm{X}}$/SFR as a function of $Z$ for both scenarios outlined above are shown in Figure \ref{fig:range}.  The horizontal bar symbols show the upper and lower bounds on $L_{\mathrm{X}}$/SFR for the stacks when using the lower or upper bounds on SFR, respectively, for galaxies which are not detected at wavelengths $\geq24 \mu$m.  The open symbols in the figure show the results for the more conservative case in which we adopt the lower and upper bounds on SFR for galaxies which are not detected at wavelengths $\geq 100\mu$m.  As can be seen in Figure \ref{fig:range}, even if we consider these extreme bounds on the $L_{\mathrm{X}}$/SFR values, the $z\approx0.2$ and $z\approx0.3$ stacks favor an anti-correlation between $L_{\mathrm{X}}$/SFR and $Z$.  The $L_{\mathrm{X}}$-SFR-$Z$ relation at $z\approx0.2$ remains steeper than at $z\approx0.3$, but at a given O/H, the $L_{\mathrm{X}}$/SFR of stacks from $z=0.2-2$ mostly remain consistent within the statistical errors (shown in Figure \ref{fig:bestfit}).  The most significant caveat introduced by the uncertainties in the SED-derived SFRs illustrated in Figure \ref{fig:range} is that, if the true SFRs are closer to the lower SFR bounds than our best estimates, the $L_{\mathrm{X}}$-SFR-$Z$ relation would be more consistent with the normalizations of the preferred \citetalias{fragos13b} and \citetalias{madau17} models.
\par
The $Z$ dependence of SFR indicators may also impact the observed slope of the $L_{\mathrm{X}}$-SFR-$Z$ relation.  We have used solar metallicity stellar population models and the Calzetti extinction curve in deriving our SFR estimates, but these assumptions may not be appropriate for some of our low-$Z$ galaxy stacks.  \citetalias{fornasini19} explored in detail the impact of using different SFR indicators on the $L_{\mathrm{X}}$/SFR values measured for the stacks of $z\sim2$ MOSDEF galaxies included in Figures \ref{fig:lxzevol}-\ref{fig:bestfit}.  Using $Z=0.004$ BC03 stellar population models and adopting the SMC extinction curve \citep{gordon03} can result in SFRs that are 0.3 dex lower than those derived using $Z=0.02$ and the Calzetti curve.  However, \citetalias{fornasini19} also found that the latter were in better agreement with H$\alpha$ derived SFRs.  If the SFRs of our low-$Z$ stacks have been overestimated, that would imply that the $L_{\mathrm{X}}$-SFR-$Z$ relation is steeper, with a slope that could more closely resemble the \citetalias{madau17} model.  As our understanding of SFR indicators at low-$Z$ and high redshift improves, it will be important to update these constraints on the $L_{\mathrm{X}}$-SFR-$Z$ relation. \par
Finally, the assumed X-ray spectrum also affects the derived $L_{\mathrm{X}}$.  Changing the spectral parameters within the range consistent with the hardness ratios of the stacks can change the measured $L_{\mathrm{X}}$ by $+/-$0.3 dex.  Adopting a different spectral model, as long as it is the same for all stacks, would affect the overall normalization of the $L_{\mathrm{X}}$-SFR-$Z$ relation, but not its slope.  If the true spectrum of the sources is harder than the $\Gamma=2$ power-law we have assumed, then our stacks would be more consistent with the mean of the highest likelihood \citetalias{fragos13b} models shown in Figure \ref{fig:lxzdep}. \par
While systematic effects may therefore impact the overall slope and normalization of the $L_{\mathrm{X}}$-SFR-$Z$ relation, it is important to note that these effects do not impact the primary conclusion that the $Z$ dependence of HMXBs is consistent across redshift and drives the observed redshift evolution of $L_{\mathrm{X}}$/SFR of star-forming galaxies.

\section{Conclusions}
\label{sec:conclusions}
We have used samples of star-forming galaxies at $z=0.1-0.9$ with optical spectra from the hCOSMOS and zCOSMOS surveys and deep \textit{Chandra} data to expand upon a previous investigation of the connection between the redshift evolution and the $Z$ dependence of HMXBs.  Using a sample of $z\sim2$ galaxies, \citetalias{fornasini19} found the first direct evidence that the redshift evolution of $L_{\mathrm{X}}$/SFR is driven by the $Z$ dependence of HMXBs.  Their conclusion relied on comparing the $L_{\mathrm{X}}$-SFR-$Z$ relation at $z\sim2$ with local measurements, which may be biased due to sample selections effects.  Thus, the goal of this work was to measure the $L_{\mathrm{X}}$-SFR-$Z$ relation at multiple redshifts, compare these results to the $z\sim2$ measurements, and provide stronger constraints for population synthesis models.  \par
Dividing our sample of $z=0.1-0.9$ galaxies by redshift and O/H, we stacked the X-ray data from the \textit{Chandra} COSMOS Legacy Survey at the galaxy locations to measure the average $L_{\mathrm{X}}$/SFR of each stack.  We find that $L_{\mathrm{X}}$/SFR and $Z$ are anti-correlated at all redshifts ($z=0.1-0.25$, $0.25-0.4$, $0.5-0.9$).  At $z\approx0.2$ and $z\approx0.3$, there are sufficient stacked detections that we can determine the best-fit of the $L_{\mathrm{X}}$-SFR-$Z$ relation independently.  A steeper anti-correlation is favored at $z\approx0.2$ compared to $z\approx0.3$, but this difference is not statistically significant. \par
When we split our sample by redshift alone, we find the mean $L_{\mathrm{X}}$/SFR values of the $z\approx0.2$, 0.3, and 0.7 samples are significantly lower than the mean $L_{\mathrm{X}}$/SFR of the $z\sim2$ sample from \citetalias{fornasini19}.  This result is consistent with previous measurements of the redshift evolution of HMXB emission.  Comparing the $L_{\mathrm{X}}$-SFR-$Z$ values of the $z\sim2$ stacks with our lower redshift stacks at similar O/H, we find very good agreement.  These results provide the strongest, direct evidence that the observed redshift evolution of $L_{\mathrm{X}}$/SFR is driven by the $Z$ dependence of HMXBs. \par
Combining all the $z>0$ stacks together, we find that the anti-correlation between $L_{\mathrm{X}}$/SFR and $Z$ is significant at $>99.9$\% confidence.  Parametrizing the $L_{\mathrm{X}}$-SFR-$Z$ relation as a power-law, we find that the best-fitting parameters are within the range measured for galaxies at $z=0$.  Comparing the $z>0$ stacked measurements with population synthesis models, we find they are in good agreement with the lower bound of the \citetalias{fragos13b} models, which is 0.15 dex lower than the mean of their six highest likelihood models.  However, systematic uncertainties associated with the X-ray spectrum and poor constraints in the MIR-FIR bands for a significant fraction of the galaxy sample could bring our results in better agreement with the preferred \citetalias{fragos13b} and \citetalias{madau17} models.  We also note that the intrinsic $L_{\mathrm{X}}$-SFR-$Z$ may be steeper than that observed due to  systematic effects arising from AGN contamination and uncertainties in the variation of SFR indicators with $Z$.  However, these systematic effects do not impact our conclusion that the $Z$ dependence of HMXBs drives the redshift evolution of $L_{\mathrm{X}}$/SFR. \par
Over the next two decades, order of magnitude improvements in studies of the $L_{\mathrm{X}}$-SFR-$Z$ relation will be enabled by the larger samples of galaxies with $Z$ and H$\alpha$ SFR measurements provided by 30-meter class optical/infrared telescopes, improved measurements of the IR SFRs of galaxies using \textit{JWST} and future FIR telescopes, and the larger and deeper X-ray surveys that will be performed by the next generation of X-ray telescopes.  \textit{ATHENA} will be capable of individually detecting XRB-dominated galaxies out to $z\sim1$ \citep{nandra13}, and the \textit{Lynx X-ray Observatory} would push this limit to $z\sim10$ \citep{vikhlinin}.  These X-ray telescopes would also allow more complete surveys of HMXB populations in nearby galaxies and improved characterization of the X-ray spectrum of HMXBs as a function of $Z$ (\citealt{basu19}; \citealt{zezas19}).  Better constraints on the $L_{\mathrm{X}}$-SFR-$Z$ will inform models of stellar evolution, including the progenitor pathways of gravitational wave sources, and improve estimates of the impact of X-ray binaries to the heating of the intergalactic medium during reionization.

\section*{Acknowledgements}

We thank J. H. Zahid, C. Maier, and G. Cresci for providing O/H measurements for galaxies from the hCOSMOS and zCOSMOS surveys.  We are grateful to B. Lehmer for providing helpful feedback that improved the quality and clarity of this paper during peer review.  We also thank I. Damjanov for helpful conversations about the hCOSMOS survey and using hCOSMOS data.  F. M. F. acknowledges support from the National Aeronautics and Space Administration through Chandra Award Number AR7-18009X and AR8-19010X issued by the Chandra X-ray Center, which is operated by the Smithsonian Astrophysical Observatory for and on behalf of the National Aeronautics Space Administration under contract NAS8-03060.  The development of CSTACK, which was used in this work, is supported by UNAM-DGAPA PAPIIT IN104216 and CONACyT 252531.  The scientific results reported in this article are based on observations made by the \textit{Chandra X-ray Observatory}, \textit{GALEX}, the \textit{Herschel Space Observatory}, and the \textit{Spitzer Space Observatory}.  This study also made use of data obtained at the MMT Observatory, a joint venture of the Smithsonian Institution and the University of Arizona, the Subaru Telescope, which is operated by the National Astronomical Observatory of Japan, the Canada-France-Hawaii Telescope (CFHT), which is operated by the National Research Council (NRC) of Canada, the Institut National des Science de l'Univers of the Centre National de la Recherche Scientifique (CNRS) of France, and the University of Hawaii, and ESO Telescopes at the La Silla Paranal Observatory under ESO programme ID 179.A-2005.   We extend special thanks to those of Hawaiian ancestry, on whose sacred mountain we are privileged to be guests.

\textit{Software}: CIAO \citep{fruscione06}, BEHR \citep{park06}, CSTACK \citep{miyaji08}

\textit{Facilities}: Chandra X-ray Observatory, GALEX, Herschel Space Observatory, Spitzer Space Telescope, MMT Observatory, Subaru Telescope


\bibliographystyle{mnras}
\bibliography{refs}





\bsp	
\label{lastpage}
\end{document}